# Magnetic skyrmion spectrum under voltage excitation and its linear modulation

Xing Chen[1], Wang Kang[1,*], Daoqian Zhu[1], Na Lei[1], Xichao Zhang[2], Yan Zhou[2], Youguang Zhang[1], Weisheng Zhao[1,*]

[1]Fert Beijing Institute, BDBC, and School of Microelectronics, Beihang University, Beijing 100191, China

[2]School of Science and Engineering, The Chinese University of Hong Kong, Shenzhen, Guangdong 518172, China

*E-mails: wang.kang@buaa.edu.cn and weisheng.zhao@buaa.edu.cn

# ABSTRACT

Magnetic skyrmions are topological quasiparticles with great potential for applications in future information storage and processing devices because of their nanoscale size, high stability, and large velocity. Recently, the high-frequency properties of skyrmions have been explored for magnon nanodevices. Here we systematically study the dynamics of an isolated skyrmion under voltage excitation through the voltage-controlled magnetic anisotropy effect in a circular thin film. A theoretical model considering the demagnetization energy, which has often been neglected or treated superficially in previous skyrmion research but is demonstrated to have importance in determining the skyrmion dynamic state, is developed. With our model, the periodic oscillation of the skyrmion radius can be solved numerically with similar precision compared to micromgnetic simulations, and the characteristic frequency ($f_c$) of the skyrmion breathing can be determined analytically with greater precision than previous studies. Furthermore, we find that the breathing skyrmion can be analogized as a modulator by investigating its linear modulation functionality under sinusoidal-form voltage excitation. Different from the conventional modulation system with complex CMOS circuits, this skyrmion "modulator" device integrates with both the modulation and carrier wave generation functionality, thus showing greater convenience and efficiency in applications. Our findings can provide useful guidance for both theoretical and experimental skyrmions research as well as the development of skyrmion-based magnonic devices with significant potential applicability in future communication system.

## I. Introduction

Magnetic skyrmions are topologically nontrivial, particle-like spin textures, which have been experimentally observed in B20-type bulk materials or ultrathin films exhibiting the Dzyaloshinskii–Moriya interaction (DMI) [1-3]. They have been extensively studied recently for potential applications in future information magnetic storage [4-6] and processing devices [7-9] because of their intrinsic properties of nanoscale size, extremely low depinning current density, and high motion velocity [10-12]. Detailed fundamental research is necessary for the development of devices that can exploit all these beneficial properties of skyrmions. An important component is the understanding of their dynamical excitations to utilize their characteristic properties and to manipulate them more efficiently [13-19]. Skyrmion breathing modes, in which the core of the swirling spin structure expands and compresses periodically over time, were first studied by micromagnetic simulations (MS) in skyrmion lattices [20] and then investigated experimentally in helimagnetic insulators [21]. However, theoretical study of skyrmion breathing is rare. For example, the properties of magnon modes localized on a ferromagnetic skyrmion were studied in Ref. [22], but the analytical result for the breathing-mode frequency of an isolated skyrmion was not validated. Ref. [16] derived and identified the precession frequency of a skyrmion but the DMI energy contribution was excluded. These studies neglected or imprecisely treated one of the most important energy contributions, the demagnetization energy (DE), or the stray field energy [16,22,23], as this energy term is difficult to treat analytically because of its nonlocal nature.

Recently, pure electric field or voltage application has been proposed as an energy-efficient method to manipulate magnetism; it is very promising for the development of skyrmion-based devices [24-26]. Extensive research on the voltage-controlled magnetic anisotropy (VCMA) effect has promoted deeper understanding of the underlying physical mechanisms as well as realizations of related skyrmionic device applications [27-30].

In this work, we study the dynamics of an isolated skyrmion under voltage excitation in a thin film with a large radius by MS and develop a theoretical model for determining the periodic oscillation of the skyrmion radius and the characteristic frequency ($f_c$) of the skyrmion breathing mode. We compare the corresponding results in the absence and presence of the DE, which is demonstrated to have importance in determining the dynamic state of a skyrmion. Our analytical result demonstrates greater accuracy and robustness compared to those from previous studies [16,22]. Moreover, we find that the oscillatory skyrmion embraces the properties of the linear modulation, i.e., amplitude modulation (AM) function, similar to a conventional modulator but with higher efficiency and convenience. Our results could offer evidence and guidance for skyrmion-related theories and experiments as well as for the development of skyrmionic devices in future communication system.

## II. Results

### A. Theoretical Model

We consider the case of a chiral magnetic skyrmion in the center of a circular ferromagnetic (FM) thin film with a large radius and perpendicular magnetic anisotropy (PMA). The skyrmion can be excited to the radially symmetrical magnon mode, or the so-called breathing mode, via the VCMA effect under an applied time-varying voltage, e.g., a sinc pulse voltage on the electrode gate, as illustrated in Fig. 1(a). In our configuration, we consider four contributions to the total energy of the system $\sigma$,

$$\sigma = t_f \iint \left[ A\varepsilon_{\text{ex}} + D\varepsilon_{\text{DM}} + K_u(1-m_z^2) + \varepsilon_{\text{d}} \right] dS, \qquad (1)$$

where $\varepsilon_{ex} = |\nabla \boldsymbol{m}|^2$ with an exchange constant $A$, and $\varepsilon_{DM} = m_z \nabla \cdot \boldsymbol{m} - \boldsymbol{m} \cdot \nabla m_z$ with an interfacial DMI coefficient $D$. The third term of the integrand is the PMA energy with constant $K_u$. The last term $\varepsilon_d$ is the DE density. Here, $\boldsymbol{m}$ is the unit magnetization vector, $m_z = \boldsymbol{m} \cdot \hat{\boldsymbol{z}}$ is the magnetization component normal to the surface of the film, $t_f$ is the thickness of the FM layer, and the integration is performed over the whole area of the FM layer. In the spherical angular parametrization, where $\boldsymbol{m} = (\sin\theta\cos\phi, \sin\theta\sin\phi, \cos\theta)$, the magnetization dynamics are well described by the Landau–Lifshitz equations,

$$\sin\theta\theta' = -\frac{\gamma}{M_s}\frac{\partial \varepsilon}{\partial \phi} - \alpha\sin^2\theta\phi', \quad \sin\theta\phi' = \frac{\gamma}{M_s}\frac{\partial \varepsilon}{\partial \theta} + \alpha\theta', \tag{2}$$

where prime denotes the derivation of the indicated variable with respect to time. $\varepsilon$, $\gamma$, and $\alpha$ represent the total energy density, the gyromagnetic constant, and the Gilbert damping parameter, respectively.

In our theoretical model, the breathing mode of a skyrmion is described by considering the time-dependent skyrmion radius $R(t)$ and angle $\varphi(t)$, which is the azimuthal angle of the magnetization $\boldsymbol{m}$ relative to the radial direction [see Fig. 1(b)], near their equilibrium positions $R(t) = R_s$ and $\varphi(t) = 0$. Here, $\varphi(t)$ is assumed to remain consistent along the radial direction of the skyrmion. Substituting the ansatz of the skyrmion profile [31-34] [see Fig. 1(d)] into Eq. (2), we derive the following coupled differential equations (see Appendix B),

$$\begin{aligned}\varphi' &= \alpha/\Delta\, R' + \frac{\gamma\Delta}{M_s} G_1(\varphi, R),\\ R' &= -\alpha\Delta\varphi' - \frac{\gamma\Delta}{M_s} G_2(\varphi, R),\end{aligned} \tag{3}$$

where the functions $G_1(\varphi, R)$ and $G_2(\varphi, R)$ are related to the total energy of the system (see Appendix B and C). Here, domain wall width $\Delta$ is defined as $\Delta = \sqrt{A/K_{eff}}$ with $K_{eff} = K_u - \mu_0 M_s^2/2$. Eq. (3) captures the properties of skyrmion breathing behavior more concisely and clearly than micromagnetic modeling can.

In the following, the breathing dynamics of a skyrmion is studied by MS and analyzed by using our theoretical model.

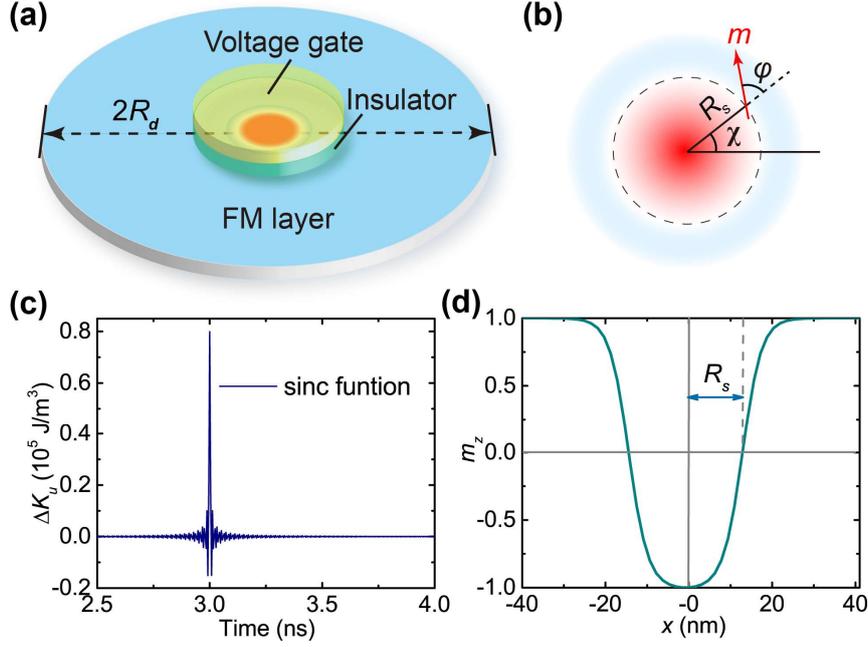

FIG. 1. Schematic of the model. (a) Schematic structure of the model. A voltage is applied on the electrode gate through an insulating layer to change the PMA of the FM layer. (b) The coordinates used in the theory. $R_s$, $\varphi$, and $\chi$ denote the equilibrium skyrmion radius, the azimuthal angle of the magnetization $m$ relative to the radial direction, and the (real space) polar angle, respectively. Here, $\phi = \chi + \varphi$. (c) Variation of the PMA in the form of the sinc function in the FM layer. (d) Illustration of the circular domain wall ansatz. The plot shows the normalized perpendicular magnetization $m_z$ as a function of position $x$ along the diameter of the skyrmion and defines the skyrmion radius $R_s$.

## B. Characteristics of breathing dynamics of a skyrmion.

Based on the modeling above, MS are performed in a circular thin film with thickness $t_f = 1$ nm and radius $R_d = 500$ nm, which is sufficiently large to avoid boundary effects. By applying a sinc pulse voltage to the electrode gate with radius $R_e = 40$ nm, the increment of the PMA of the FM layer beneath the electrode gate varies accordingly, as, $\Delta K_u = K_0 \sin[2\pi f_0 (t-t_0)]/[2\pi(t-t_0)]$ with $f_0 = 100$ GHz, $t_0 = 1$ ns, and $K_0$ being the amplitude of the excitation [see Fig. 1(c)], because of the VCMA effect [35-38] (see Appendix A for the simulation details). The breathing mode is clearly observed under the excitation [see Fig. 2(a)]. Thus, we can determine $f_c$ of the breathing mode by performing the fast Fourier transform (FFT) on the skyrmion radius $R(t)$ [see Appendix A for FFT calculations]. Normally, a greater strength causes more prominent oscillation of $R(t)$, which also suppresses $f_c$ slightly [see Fig. 2(b)] because of the increased relaxation distance ($a_0 = |R(t)_{max} - R_s|$, defined as the maximum derivation from the equilibrium radius $R_s$). In the MS study, we focus on the case of small oscillations, where $a_0$ is within 4 nm.

This breathing mode of the skyrmion closely relies on the material properties. For instance, MS results of $f_c$ and $R_s$ under different $D$ and $K_u$ are respectively shown in Fig. 2(c) and Fig. 2(d). Significantly, $f_c$ is negatively correlated with $R_s$. For quantitative analysis, we firstly numerically solve our theoretical model [Eq. (3)] by using Runge-Kutta method to obtain time varying $R(t)$, then, the equilibrium position $R_s$ can be extracted and the characteristic frequency $f_c$ can be determined by conducting FFT for $R(t)$. The results shown in Fig. 2(c) and Fig. 2(d) validate the accuracy of our model. Further, the simplified forms of the functions $G_1(\varphi, R)$ and $G_2(\varphi, R)$ (see Appendix C) can be expressed through rough approximations [39,40] for energy terms in Eq. (1) in the case of $R \gg \Delta$. Then, after the first order of approximation in Eq. (3), the corresponding static equilibrium solution for $R_s$ and the linear dynamics near the equilibrium position for $f_c$ when $a_0 \ll R_s$ can be characterized by (see Appendix E),

$$\begin{aligned} R_s &= \sqrt{C_3/C_2}, \\ f_c &= \sqrt{2C_1 C_2}/(2\pi R_s), \end{aligned} \quad (4)$$

where $C_1 = \pi \gamma D/(2M_s)$, $C_2 = \gamma/M_s (A/\Delta + K_u \Delta - \pi D/2 - k_1 \mu_0 M_s^2 \Delta)$, and $C_3 = \gamma A \Delta/M_s + k_2 \gamma \mu_0 M_s \Delta^3$ with $k_1 = 0.576$, and $k_2 = 0.158$. Here, the coefficients $k_1$ and $k_2$ are derived from the modified DE term [40,41] and the values depend on the film thickness ($t_f = 1$ nm in our case) (see Appendix C). In the limit of true two-dimensional film ($t_f \to 0$), our model converges to the case where $k_1 = 0.5$, and $k_2 = 0$. The formula results [Eq. (4)] in relation with $D$ and $K_u$ are also shown in Fig. 2(c) and Fig. 2(d) and demonstrate good agreement with the MS results, especially for a large $R_s$. More importantly, our results are more accurate in comparison with formulas from some previous studies (see Discussion).

As noted in the introduction, DE was often excluded or crudely approximated by correcting the anisotropy constant $K_u$ as an effective anisotropy $K_{eff}$, which is satisfactory in the limit of $t_f \to 0$. However, in practical situations, we find that this treatment induces some errors, especially in characterizing the dynamic properties of a skyrmion. To clarify, we compare the numerical results of $R_s$ and $f_c$ obtained by substituting the anisotropy constant with $K_{eff}$ to solve Eq. (3) with those obtained by separately calculating the DE contributions. Fig. 2 reveals that approximating the DE term by using $K_{eff}$ is sufficiently precise to evaluate the static state of a skyrmion, e.g., its $R_s$, which may be why this method has been extensively adopted in previous studies. In contrast, the approximation is not as suitable in determining the dynamical characteristics of the skyrmion, e.g., $f_c$, using this modification.

Our model is also verified when DE is excluded (see Appendix D). Fig. 2(e) and Fig. 2(f) display the results of $f_c$ and $R_s$ as a function of $D$ and $K_u$ in the absence of the DE. In this case, formula approximations for $f_c$ and $R_s$ can be readily expressed by setting the coefficients $k_1$ and $k_2$ equal to zero in Eq. (4). The numerical results of Eq. (3) using simplified form of functions $G_1(\varphi, R)$ and $G_2(\varphi, R)$ are also obtained, which show very good agreement with the formula results. However, the errors of the formula approximations for $f_c$ compared to the MS results are caused by the imprecise estimation of energy contributions related to $G_1(\varphi, R)$ and $G_2(\varphi, R)$, as opposed to the first order of approximation. In specific, the ansatz to describe the skyrmion profile, i.e., the circular domain wall ansatz, is well represented for large skyrmion sizes ($R \gg \Delta$), but exists some errors for a skyrmion with a size comparable to $\Delta$, thus yielding imprecise approximations for $G_1(\varphi, R)$ and $G_2(\varphi, R)$. Nevertheless, our formula expressions still show higher accuracy than previous studies (see Discussion).

Additionally, the skyrmion can stabilize with a relatively lower $D$ value in the presence of DE. This is easily observed by the reality that $C_2 > 0$, from which we can determine the upper limit of $D$, i.e., the critical value for $D$,

$$D_c = \frac{2}{\pi}\left(A/\Delta + K_u\Delta - k_1\mu_0 M_s^2\Delta\right). \tag{5}$$

Therefore, we can obtain the restriction of $D < 4.41 \, \mathrm{mJ/m^2}$ when DE is excluded from the system and $D < 3.78 \, \mathrm{mJ/m^2}$ when DE is included in the system for $K_u = 0.8 \, \mathrm{MJ/m^3}$ in our simulations.

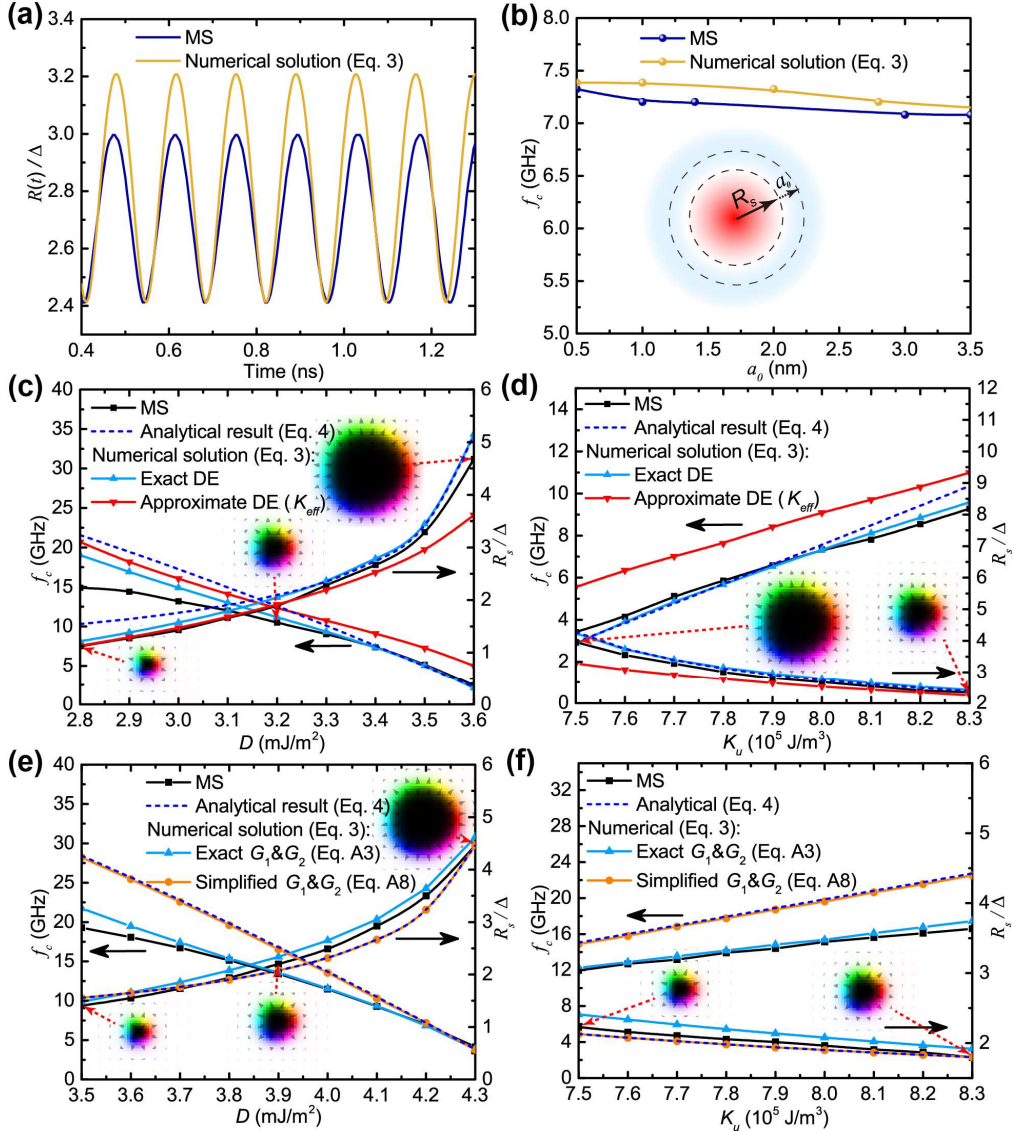

FIG. 2. Comparative analysis between the results from MS and our theoretical model. (a), (b): Comparison between the MS results (blue) and the numerical solutions (yellow) of Eq. (3) for the normalized skyrmion radius $R(t)/\Delta$ with respect to time in (a) and for $f_c$ as a function of $a_0$ in (b). Here, $D = 3.4 \, \mathrm{mJ/m^2}$ and $K_u = 0.8 \, \mathrm{MJ/m^3}$. The inset in (b) defines the amplitude of the oscillation $a_0$. (c), (d): MS results (black),

numerical solutions of Eq. (3) using the exact DE (blue), numerical solutions of Eq. (3) by including DE with an effective anisotropy ($K_{eff}$) (red) and analytical results [Eq. (4), dashed lines] for $f_c$ and $R_s$ as a function of $D$ in (c) and $K_u$ in (d). Functions $G_1(\varphi, R)$ and $G_2(\varphi, R)$ in Eq. (3) are calculated by using Eq. (A3). (e), (f): MS results (black), numerical solutions of Eq. (3) using exact $G_1(\varphi, R)$ and $G_2(\varphi, R)$ [Eq. (A3), blue], numerical solutions of Eq. (3) using simplified $G_1(\varphi, R)$ and $G_2(\varphi, R)$ [Eq. (A8), orange], and analytical results [Eq. (4), dashed lines] for $f_c$ and $R_s$ as a function of $D$ in (e) and of $K_u$ in (f) when DE is absent. Here, $K_u = 0.8 \text{ MJ}/\text{m}^3$ for (c) and (e), $D = 3.4 \text{ mJ}/\text{m}^2$ for (d) and $D = 3.8 \text{ mJ}/\text{m}^2$ for (f). Other key parameters adopted in simulations are: the exchange stiffness $A = 15 \text{ pJ}/\text{m}$, saturation magnetization $M_s = 580 \text{ kA}/\text{m}$.

### C. Breathing skyrmion as a modulator.

We have demonstrated that the breathing frequency $f_c$ of a skyrmion is an intrinsic characteristic closely associated with the material properties. To further investigate and utilize this property, it is also important to explore the skyrmion dynamics under various forms of excitation, such as single-frequency excitation. According to the solutions to skyrmion radius of our theoretical model (see Appendix A), the skyrmion breathes in hybrid frequencies of the external sine frequency $f_e$ and $f_c$ (see Fig. 3) under sine wave voltage excitation. Moreover, the relative amplitude of $f_e$ and $f_c$ in the spectrum depends on the value of $f_e$. Specifically, if $f_e < f_c$, the skyrmion breathing typically synchronizes with $f_e$, because the external driving frequency $f_e$ is dominant. Conversely, if $f_e > f_c$, the skyrmion is more likely to breath at $f_c$. This phenomenon is very similar to the forced oscillation of a harmonic oscillator: for a periodic force applied to a harmonic oscillator, the output signal is a superposition of two sine waves with the external forced frequency and the eigen frequency respectively, according to the solution of the well-known resonance differential equation. In this respect, skyrmion is analogized to a conventional "oscillator" with tunable eigen frequency. However, differently in the skyrmion "oscillator", other prominent frequencies can also be discovered in addition to the two main peaks in the frequency spectrum, which reveals the "resonant skyrmion" also functions as a "modulator."

In conventional AM technology, the frequency spectrum of message signal is shifted to a higher frequency band via a radio carrier wave for transmitting information. Typically, the frequency of message signal is much lower than the carrier frequency. In Fig. 3 (b) and (c), we display the power spectral density (PSD) of skyrmion radius (see Appendix A) corresponding to $f_e = 12$ GHz and $f_e = 3$ GHz respectively in Fig. 3(a). Here, the input excitation signal is viewed as the message signal, and the skyrmion radius variation is the output signal, which can be readily detected by the fluctuations of the average perpendicular component of the magnetization in experiment (see Appendix F for MS results). In Fig. 3(c) where $f_e < f_c$, two relative strong peaks with frequency $f_c - f_e$ and $f_c + f_e$ show the amplitude modulated signal, expressed by,

$$S(t) = S_0 \sin(2\pi f_e t) \times \sin(2\pi f_c t), \tag{6}$$

with $S_0$ being the relative amplitude. Here, two other frequency components $2f_e$ and $3f_e$ in Fig. 3(c) are harmonic frequencies of message signal, which can be depressed or eliminated by a harmonic filter. In comparison with conventional AM where the modulation of message signal is realized by using a multiplier via integrated CMOS circuits, and the carrier wave is generated by crystal oscillator, this skyrmion

"modulator" integrates with the modulation and carrier wave generation functionality, thus is more convenient and efficient in applications. For more information about the time-domain behaviors of skyrmion and comparisons between the theoretical results and the MS results, see Appendix F.

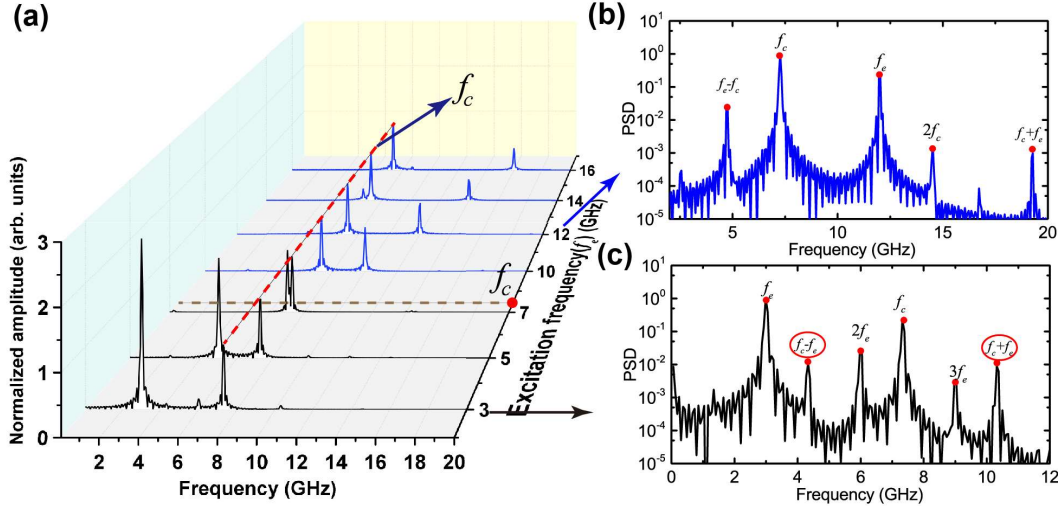

FIG. 3. Theoretical results of Fourier transform spectra of the skyrmion radius under single-frequency excitations. (a) The amplitude of the Fourier transform spectrum under each excitation frequency is normalized by the corresponding amplitude at the position of $f_c$, indicated by the red dashed line. The results are obtained by numerically solving Eq. (5). PSD of skyrmion radius at $f_e = 12$ GHz and $f_e = 3$ GHz are shown in (b) and (c), respectively. The two circles in (c) correspond to the amplitude modulated signal.

We also investigate the skyrmion dynamics under sine excitation with frequency close to the skyrmion characteristic frequency. In this case, the resonant oscillation of the skyrmion is expected to be seen. Here, the amplitude of the excitation sine wave is set to a very small value compared to those in Fig. 3(a) to avoid the wild oscillation or destruction of the skyrmion (see Appendix A). Both the numerical solutions of Eq. (3) and MS results are discussed in the following. Interestingly, instead of showing the resonant oscillation phenomenon, the variation of $R(t)/\Delta$ is packaged by a periodic wave with a lower frequency [see Fig. 4(a)]. The fluctuant range of the radius $R(t)$ is ~6 nm, despite the very low strength of the excitation signal. The fluctuations of the average perpendicular component of the magnetization $\delta m_z(t) = \langle m_z \rangle(t) - \langle m_z \rangle(0)$, where $\langle m_z \rangle(0)$ is the equilibrium state and the angular bracket denotes the spatial average of the magnetization, is also shown in Fig. 4(b). It is apparent that $\delta m_z(t)$ and $R(t)/\Delta$ show consistent patterns, which are also identified from the corresponding PSD in Fig. 4(c). In the middle of the PSD, a peak exists very near the position of the frequency $f_e$, indicating that the exact characteristic frequency $f_c$ has shifted slightly towards a smaller value because of the relatively large oscillation [see relation between $f_c$ and $a_0$ in Fig. 2(b)]. Two additional distinct peaks appearing on both sides, corresponding to the values of $f_c - f_e$ and $f_c + f_e$, is the modulated signal, similar to the case in Fig. 3(c). The package frequency in Fig. 4(a) and Fig. 4(b) is actually related to $f_c - f_e$. This phenomenon demonstrates vividly that the skyrmion itself is able to carry information by linear superposition of input signal and the excited eigen signal.

The skyrmion also shows a type of "swing" behavior. Specifically, the symmetry axis of the magnetization projecting in the *xy* plane rotates back and forth during the breathing process [see Fig. 4(d)]. This phenomenon arises from the magnetization variations along the radial direction, i.e., the angles $\varphi$ are inconsistent because of the drastic changes in the skyrmion radius. Through the analysis above, the resonant skyrmion functions as an oscillator as well as a modulator, thus is expected to achieve device miniaturization for the promising prospect in future communication system.

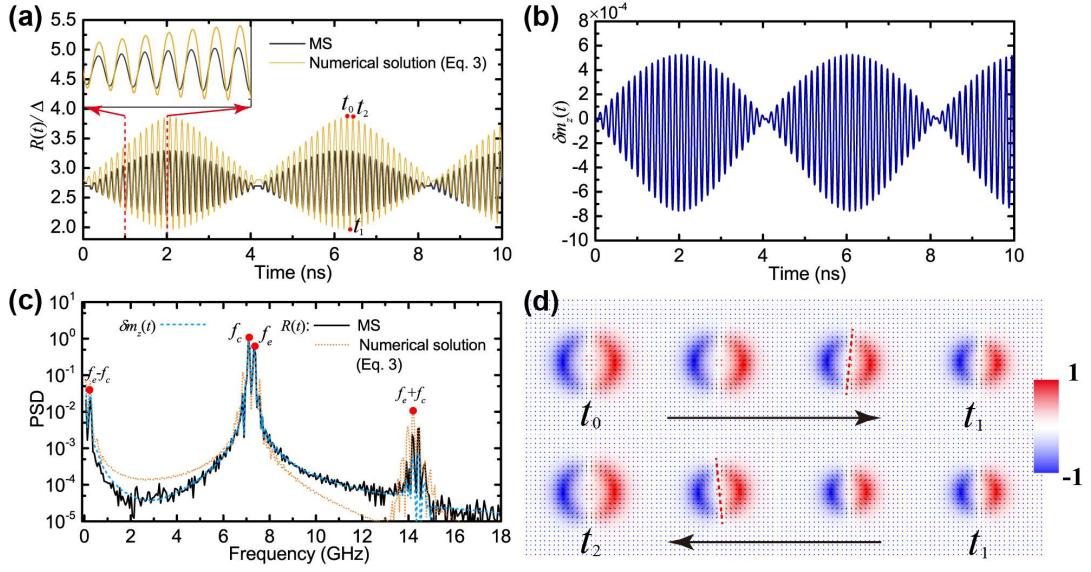

FIG. 4. Skyrmion dynamics in excitation by sine-wave voltage ($f_e = f_c = 7.342$ GHz). (a) MS results (black) and numerical solutions of Eq. (3) (yellow) for $R(t)/\Delta$. The inset shows the detailed view between 1 ns and 2 ns. (b) MS results of $\delta m_z(t)$. (c) PSD for $R(t)$ from MS (black) and numerical solutions of Eq. (3) (orange), as well as for $\delta m_z(t)$. (d) Variation of skyrmion profile (magnetization in $x$ direction) from time $t_0$ to time $t_1$ [indicated in (a)] and from time $t_1$ to time $t_2$ [indicated in (a)] in the breathing process. The red dashed line indicates the symmetry axis of the skyrmion profile, which rotates slightly compared to the equilibrium state.

### III. Discussion

Up to now, we have discussed the formula approximation for the characteristic frequency $f_c$ of the skyrmion breathing mode. For greater clarification, we further compare the results from our formula approximations with those from previous research. In Ref. [16], the normalized characteristic frequency $f_c^*$ and the skyrmion radius $R_s^*$ obey the simple relation $2\pi f_c^* R_s^* = 1$, where we obtain $f_c \propto 1/R_s^*$. This result is obtained when the DMI energy contribution is excluded and the skyrmion radius is tuned by the drive current. In Ref. [22], the normalized frequency $f_c^*$ is characterized by $f_c^* = 1/(2\pi \cdot R_s^* \sqrt{1+R_s^*}) \propto 1/R_s^{*2}$ with $R_s$ being normalized. Meanwhile, in our studies, derived from Eq. (4), the relation between $f_c$ and $R_s$ is expressed as $f_c = \gamma A/(\pi \Delta^2 M_s)\sqrt{1/(R_s/\Delta)^4 - 0.5/(R_s/\Delta)^6}$, from which we determine $f_c \propto \sqrt{1/R_s^{*4} - 1/(2R_s^{*6})}$. If the skyrmion radius is assumed to be sufficiently large, we obtain $f_c \propto 1/R_s^{*2}$, which is the same with that in Ref. [22]. Note that the normalizations used in different studies are different. A comparison for the case of DE inclusion is also shown in Fig. 5(b). In brief, the results in Fig. 5 demonstrate the relatively higher accuracy and generality of our formula approximation.

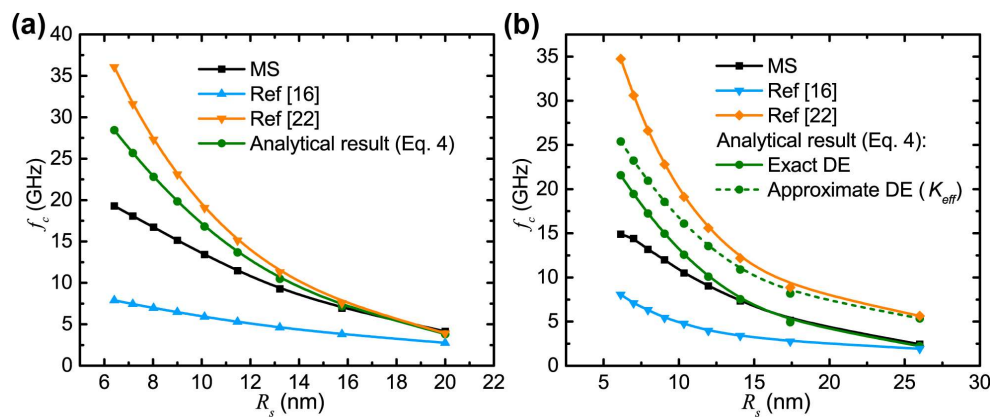

FIG. 5. Comparisons of $f_c$ between the results from our analytical results and those from previous literatures. (a) Calculations of $f_c$ as a function of $R_s$ by using our analytical results (green) of Eq. (4), formulas from Ref. 16 (blue) and Ref. 21 (orange) in the absence of DE in (a) and in the presence of DE in (b), respectively. The black curves show the MS results. The dashed line in (b) indicates the analytical results of Eq. (4), where DE is considered by using $K_{eff}$. Note that $R_s$ is varied by changing the DMI constant, corresponding to Fig. 2(c) and Fig. 2(e).

## IV. Conclusion

To conclude, we have studied skyrmion dynamics under voltage excitation by using the VCMA effect. The skyrmion breathing behaviors were systematically investigated via both MS and our theoretical model. Our model not only provides very exact numerical solutions for the time-dependent skyrmion radius regardless of the form of excitation, but also yields corresponding analytical solutions, which are more accurate and robust than those from previous studies. In addition, the AM functionality of a skyrmion has also been investigated, which shows great applicability potential in future communication systems, and may provide guidance for the design of skyrmion-based high-frequency magnonic devices.

## ACKNOWLEDGEMENTS

The authors gratefully acknowledge the National Natural Science Foundation of China (Grant No. 61871008, 61571023, 61627813, 11574018), the National Key Technology Program of China (2017ZX01032101) for their financial support of this work. X.Z. acknowledges the support by the Presidential Postdoctoral Fellowship of The Chinese University of Hong Kong, Shenzhen (CUHKSZ). Y.Z. acknowledges the support by the President's Fund of CUHKSZ, the National Natural Science Foundation of China (Grant No. 11574137), and Shenzhen Fundamental Research Fund (Grant Nos. JCYJ20160331164412545 and JCYJ20170410171958839).

## APPENDIX A: SIMULATION METHODS

Micromagnetic simulations (MS) are performed by using the graphics-processing-unit-based tool MuMax3 [42]. The default mesh size of 1.5625

nm× 1.5625 nm× 1 nm is used in our simulations. We adopted the following material parameters in our simulations: the exchange stiffness $A=15$ pJ/m, saturation magnetization $M_s = 580$ kA/m and default PMA constant of the FM layer $K_u = 0.8$ MJ/m$^3$ [8,11,16]. In addition, we set the VCMA coefficient $\xi$ as 100 fJ·V$^{-1}$m$^{-1}$ based on recent experiments [35-38]. Here the typical thickness of the insulating layer is 1 nm. Under these conditions, with an applied voltage of 0.1 V (an electric field of 0.1 V/nm), PMA constant in FM layer will change 10 kJ/m$^3$, which is about 1.25% of the PMA change. In the case where the DE is not included, the maximum amplitude of the variation of the PMA of the FM layer in excitation with the sinc voltage is $K_0 = 0.1$ MJ/m$^3$, which corresponds to an applied voltage of 1 V. Considering the effect of the demagnetization field in decreasing the degree of varying the magnetization in the $z$ axis, the maximum amplitude of the excitation pulse $K_0 = 0.08$ MJ/m$^3$ (corresponding to 0.8 V) is consequently used to ensure the small oscillations of the skyrmion in the case where DE is included.

In terms of the single frequency excitation case, based on our analytical model, we firstly calculated the variations of the skyrmion radius in excitation with voltages with different $f_e$ by solving Eq. (5), where $K_u$ is set as $K_u = K_0 \sin(2\pi f_e t)$ with $K_0 = 0.01$ MJ/m$^3$ when $f_e = 3$ GHz ($<f_c$), 12 GHz ($>f_c$) and $K_0 = 0.001$ MJ/m$^3$ when $f_e = f_c = 7.342$ GHz, then the Fourier transform spectrum is obtained by performing the Fast Fourier transforms.

In our simulations, the variation of the skyrmion radius $R(t)$ and the average magnetization $\langle m_z \rangle(t)$ are extracted for over 20 periods and the sampling frequency $f_s$ is 250 GHz, which satisfies the Nyquist sampling theorem ($2f_0 < f_s$). Then, the Fast Fourier transforms (FFT) is performed as, $F(m) = \sum_{t_n} (s(t_n) - s(t_0)) \cdot W_N^{mn}$, where $m = 0, 1, \cdots, N-1$, $W_N = e^{-j2\pi/N}$ with $N$ being the transformation point and $s(t_n) = R(t_n)$ or $\langle m_z \rangle(t_n)$. Thus, we can obtain the Fourier transform spectrum $F(f)$, from which $f_c$ is determined by the position of the maximum peak. The power spectrum density (PSD) is computed by using the normalized spectrum $F^*(f) = F(f)/F(f_c)$, as, $\text{PSD} = \log |F^*(f)|^2$.

## APPENDIX B: DERIVATION OF THE THEORETICAL MODEL

For a skyrmion centered at a point $r = 0$, the polar and azimuthal angles $\theta$ and $\phi$ of the magnetization at $r$ can be described by [10],

$$\theta = \theta(r), \quad \phi = Q_v \Phi + Q_h, \tag{A1}$$

with $Q_v$ being the vorticity number ($Q_v = 1$ for a skyrmion, $Q_v = -1$ for an antiskyrmion) and $Q_h$ being the helicity number ($Q_h = 0$ or $\pi$ for the Néel-type skyrmion and $Q_h = \pm\pi/2$ for the Bloch-type skyrmion). In our simulations, a Néel-type skyrmion is adopted. For a free skyrmion in a nanostructure, the skyrmion profile is described by the circular domain wall ansatz [31-34] [see Fig. 1(d)],

$$\cos\theta = \tanh\left(\frac{r - R_s}{\Delta}\right), \Phi = \chi + \varphi, \tag{A2}$$

which has been experimentally used to determine the skyrmion profile. Here, $R_s$, $\varphi$ and $\Delta$ denote the equlibrium skyrmion radius, the azimuthal angle of the magnetization $m$ relative to the radial direction and domain wall width, respectively. It should be noted that the radial coordinate $r$ and the (real space) polar angle $\chi$ correspond to the polar coordinates [see Fig. 1(b)]. The excitation of the breathing mode of skyrmion can be described by considering the time-dependent skyrmion radius $R(t)$ and angle $\varphi(t)$, which is assumed to keep consistent along the radial direction of skyrmion, in the vicinity of their equilibrium positions $R(t) = R_s$ and $\varphi(t) = 0$. Inserting the time-varying ansatz of Eq. (A2) into Eq. (2), we consequently obtain the coupled differential equations after a 2D integration over the whole FM layer, as displayed in Eq. (3), where the functions $G_1(\varphi, R)$ and $G_2(\varphi, R)$ lead the following forms,

$$G_1(\varphi, R) = \frac{\partial \sigma / \partial R}{t_f \int_0^\infty \sin^2 \theta r dr},$$

$$G_2(\varphi, R) = \frac{\partial \sigma / \partial \varphi}{t_f \int_0^\infty \sin^2 \theta r dr}. \tag{A3}$$

Here, the total energy $\sigma = \sigma(R, \varphi) = t_f \iint \varepsilon dS = t_f \iint \left[ A\varepsilon_{\text{ex}} + D\varepsilon_{\text{DM}} + K_u(1 - m_z^2) + \varepsilon_d \right] dS$, which can be numerically integrated by substituting the skyrmion profile ansatz [Eq. (A2)] into the integrand. We noticed other method has also been developed to derive the skyrmion breathing behavior, instead, based on Hamiltonian mechanics [23], which verifies our results.

**APPENDIX C: CALCULATION OF FUNCTIONS $G_1(\varphi, R)$ AND $G_2(\varphi, R)$**

Calculations of the functions $G_1(\varphi, R)$ and $G_2(\varphi, R)$ involve the integration of the total energy density $\varepsilon$ over the system, which is difficult to express analytically. However, it is possible to approximate it by separate integration of each energy contribution in the case of $R \gg \Delta$ as [39-40],

$$E_{\text{ex}} = t_f A \iint \varepsilon_{\text{ex}} dS \approx 4\pi A t_f (R/\Delta + \Delta/R), \tag{A4}$$

$$E_{\text{DM}} = t_f D \iint \varepsilon_{\text{DM}} dS \approx 4\pi D t_f (-\pi R \cos\varphi/2), \tag{A5}$$

$$E_{\text{an}} = t_f K_u \iint (1 - m_z^2) dS \approx 4\pi K_u t_f \Delta R, \tag{A6}$$

$$E_d = E_{d,s} + E_{d,v} \approx -4\pi\mu_0 M_s^2 t_f \Delta^2 (k_1 R/\Delta - k_2 \Delta/R). \tag{A7}$$

Note that the stray field energy $E_d$ consists of surface stray field energy ($E_{d,s}$) and volume stray field energy ($E_{d,v}$). In the limit of ultra-thin films, surface stray field acts as an effective anisotropy of strength $-1/2\mu_0 M_s^2$ and volume stray field is negligible. The full stray field integral can be found in Ref. [40, 43-45]. In our case, only $E_{d,s}$ is taken into consideration for simplicity when simplifying functions $G_1(\varphi, R)$ and $G_2(\varphi, R)$. Specifically, $E_{d,s} = -4\pi\mu_0 M_s^2 t_f \Delta^2 I_s$ involves highly non-trivial integrals in $I_s$ [40], which relies on skyrmion radius $R$ and film thickness $t_f$. The coefficients $k_1$ and $k_2$ in Eq. (A7) is obtained by fitting the integrals of $I_s$ when $t_f = 1$ nm. Therefore, the functions $G_1(\varphi, R)$ and $G_2(\varphi, R)$ can be expressed as,

$$\begin{aligned} G_1(\varphi, R) &= A\left(\frac{1}{\Delta^2 R} - \frac{1}{R^3}\right) - \frac{\pi D}{2\Delta}\frac{\cos\varphi}{R} + \frac{K_u}{R} - \mu_0 M_s^2 \Delta^2 \left(\frac{k_1}{\Delta^2 R} + \frac{k_2}{R^3}\right), \\ G_2(\varphi, R) &= \frac{\pi D}{\Delta}\sin\varphi, \end{aligned} \tag{A8}$$

where the width $\Delta = \sqrt{A/K_{\text{eff}}}$ and $\mu_0$ is the magnetic vacuum permeability.

**APPENDIX D: NUMERICAL SOLUTION OF EQ. (3) IN THE ABSENCE OF DE**

In Fig. 2(a) in the main text, we show the accuracy of numerical solutions of Eq. (3)

via numerically integrating each energy term to calculate functions $G_1(\varphi, R)$ and $G_2(\varphi, R)$. Here, we display the corresponding results in the absence of DE [see Fig. 6(a)]. Besides, a larger oscillation amplitude $a_0$ will drop down $f_c$ a little [see Fig. 6(b)], similar to Fig. 2(b) in the main text.

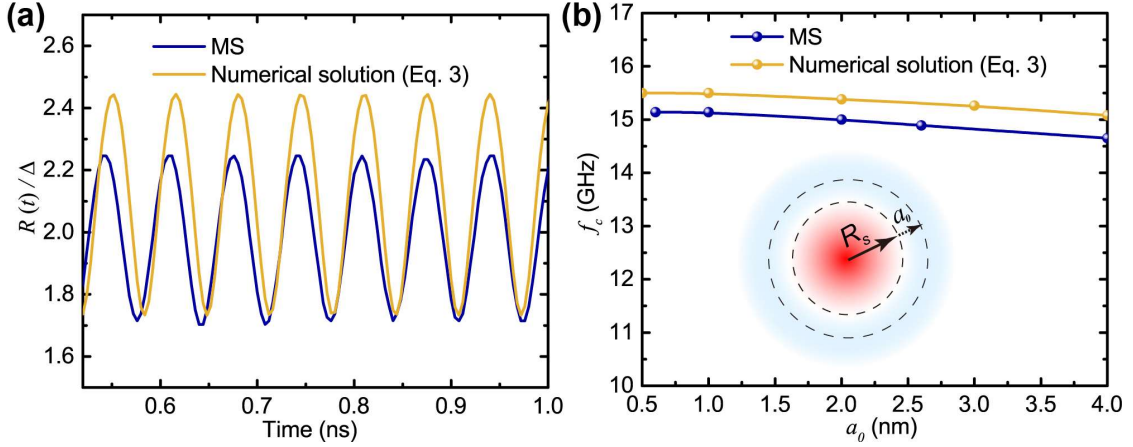

FIG. 6. Comparison between the MS results (blue) and the numerical solutions (yellow) of Eq. (3) for the normalized skyrmion radius $R(t)/\Delta$ with respect to time in (a) and for $f_c$ as a function of $a_0$ in (b) in the absence of DE. Here, $D = 3.8 \text{ mJ}/\text{m}^2$ and $K_u = 0.8 \text{ MJ}/\text{m}^3$.

## APPENDIX E: ANALYTICAL SOLUTION OF QUATION (3)

To determine the analytical solution of Eq. (3), we consider an approximate solution with the form of $R(t) = R_s + a_0 \sin(2\pi f t)$ based on the oscillation behaviour of skyrmion. Taking this form of $R(t)$ and the Eq. (A8) into the Eq. (3), we are able to determine $R_s$ and $f_c$ after the first order of approximation as,

$$R_s = \sqrt{C_3/C_2},$$
$$f_c = \frac{1}{2\pi}\sqrt{\frac{2}{R_s^4}\left(\sqrt{C_2^2 a_0^4 + R_s^4 C_1^2 C_2^2} - C_2^2 a_0^2\right)}, \quad (A9)$$

where $C_1 = \pi\gamma D/(2M_s)$, $C_2 = \gamma/M_s\,(A/\Delta + K_u\Delta - \pi D/2 - k_1\mu_0 M_s^2\Delta)$, and $C_3 = \gamma A\Delta/M_s + k_2\gamma\mu_0 M_s\Delta^3$ with $k_1 = 0.576$ and $k_2 = 0.158$. If the oscillation amplitude $a_0 \to 0$, we can obtain the Eq. (4) in the main text. Actually, $f_c$ drops down a little if $a_0$ is set as a positive value, which shows accordance with the result in Fig. 1(b) in the main text and Fig. 6(b) (see Fig. 7).

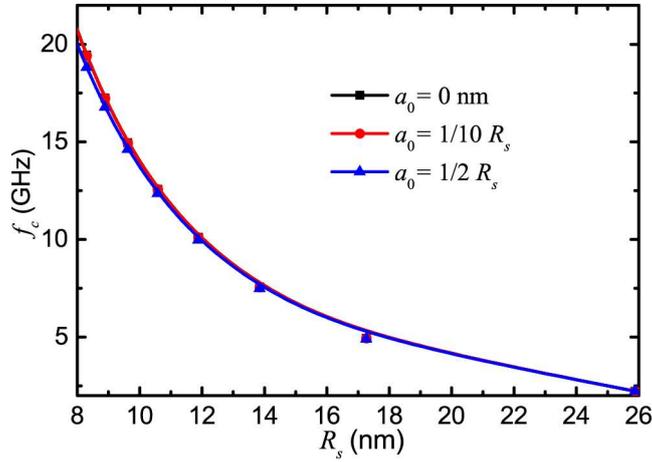

FIG. 7. $f_c$ as a function of $R_s$ for various $a_0$ by using Eq. (A9). Here, $R_s$ is varied by changing the DMI constant from $D = 2.8$ mJ/m$^2$ to $D = 3.6$ mJ/m$^2$, the range of which is the same with that in Fig. 2(c) in the main text.

**APPENDIX F: SKYRMION DYNAMICS UNDER SINE WAVE EXCITATION**

In the main text, we have discussed in detail the skyrmion dynamics under a voltage with frequency $f_e = f_c = 7.342$ GHz. Here, we examine the behaviours of skyrmion when $f_e$ is set as 3 GHz ($<f_c$) and 12 GHz ($>f_c$) respectively.

In Fig. 8, we show the fluctuations of $\delta m_z(t)$ and $R(t)/\Delta$ as well as the corresponding PSD. It is found that our theoretical results show good agreement with MS results. Analyzing from the PSD, the overall response signal $S(t)$ of the skyrmion dynamics [$\delta m_z(t)$ and $R(t)/\Delta$] can be expressed as a superposition of the excitation signal ($f_e$), characteristic signal ($f_c$) and the multiplication signal,

$$S(t) = S_1 \sin(2\pi f_e t)\left(1 + S_2 \sin(2\pi f_c t)\right) + S_3 \sin(2\pi f_e t), \tag{A10}$$

where $S_1, S_2$ and $S_3$ are relative amplitudes. In the case where $f_e = 3$ GHz ($<f_c$), the response signal contains the modulated signal of the excitation signal with frequency $f_e$ by the characteristic signal with frequency $f_c$, leading to the two distinct peaks with frequencies $f_c - f_e$ and $f_c + f_e$ in the PSD. Here the two other frequency components $2f_e$ and $3f_e$ in the PSD are harmonic frequencies of message signal, which can be depressed or eliminated by a harmonic filter. Case is similar when $f_e = 12$ GHz ($>f_c$), except that the role of carrier wave and message wave replace with each other.

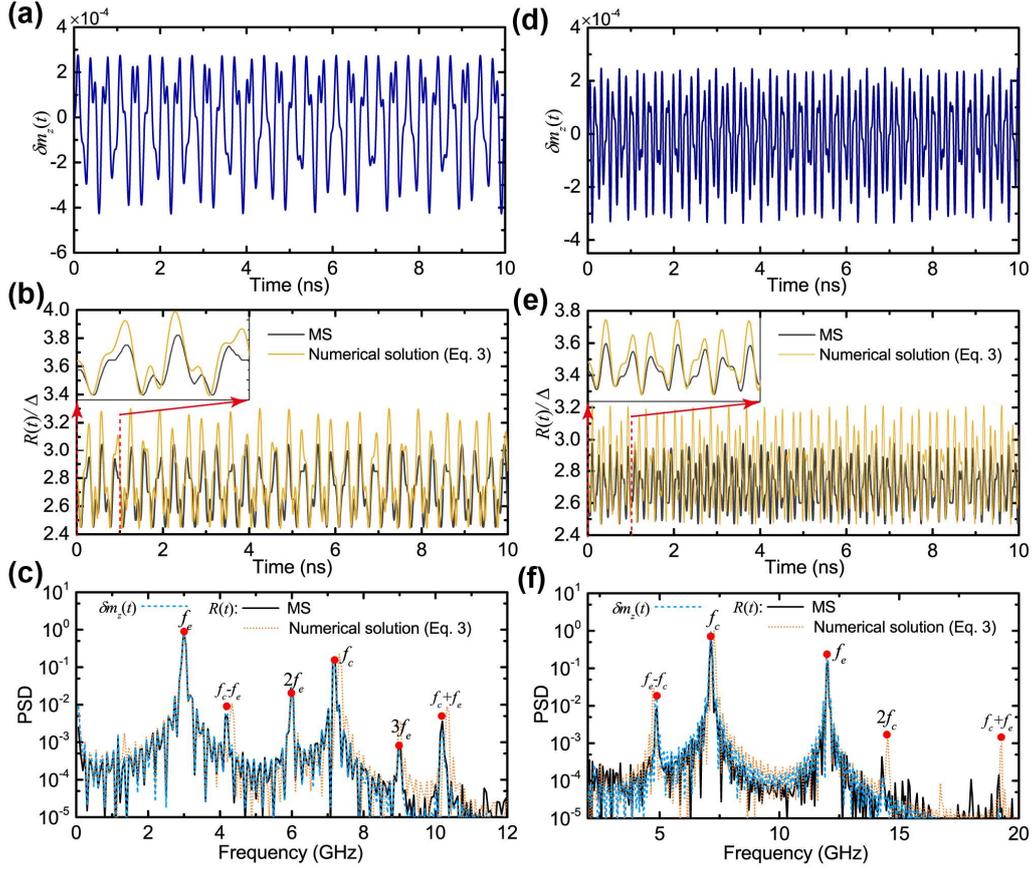

FIG. 8. Skyrmion dynamics under sine-wave excitations ( $f_e = 3$ GHz ($<f_c$), 12 GHz ($>f_c$) ). MS results of the fluctuations of $\delta m_z(t)$ in (a) ( $f_e = 3$ GHz ) and (d) ( $f_e = 12$ GHz ). MS results (black) and theoretical results (yellow) of the variation of $R(t)/\Delta$ in (b) ( $f_e = 3$ GHz ) and (e) ( $f_e = 12$ GHz ). PSD of the variation of $R(t)$ from MS (black) and theoretical results (orange), as well as $\delta m_z(t)$ (blue) in (c) ( $f_e = 3$ GHz ) and (f) ( $f_e = 12$ GHz ). The insets in (b) and (e) show the detailed views between 0 ns and 1 ns.